\documentclass[twocolumn,secnumarabic,amssymb, nobibnotes, aps, prd]{revtex4-1}

\usepackage{graphicx}

\begin{document}

\title{Scaling Behavior of Quasi-One-Dimensional Vortex Avalanches in Superconducting Films}
\author{A. J. Qviller}
\affiliation{nSolution AS, Maries gt. 6, 0368 Oslo, Norway}
\author{T. Qureishy}
\affiliation{Department of Physics, University of Oslo, P. O. Box 1048 Blindern, 0316 Oslo, Norway}
\author{Y. Xu}
\affiliation{The Key Laboratory of Advanced Functional Materials, Ministry of Education, Beijing University of Technology,
Beijing 100022, China}
\affiliation{Department of Energy Conversion and Storage, Technical University of Denmark, Roskilde 4000, Denmark}
\author{H. Suo}
\affiliation{The Key Laboratory of Advanced Functional Materials, Ministry of Education, Beijing University of Technology,
Beijing 100022, China}
\author{P. B. Mozhaev}
\affiliation{Institute of Physics and Technology of the Russian Academy of Sciences, Moscow, 117218, Russia}
\author{J. B. Hansen}
\affiliation{Department of Physics, Technical University of Denmark, Kongens Lyngby, DK-2800, Denmark}
\author{J. I. Vestg\aa rden}
\affiliation{Department of Physics, University of Oslo, P. O. Box 1048 Blindern, 0316 Oslo, Norway}
\affiliation{Norwegian Defence Research Establishment (FFI), P. O. Box 25, 2027 Kjeller, Norway}
\author{T. H. Johansen}
\affiliation{Department of Physics, University of Oslo, P. O. Box 1048 Blindern, 0316 Oslo, Norway}
\affiliation{Institute for Superconducting and Electronic Materials, University of Wollongong, Northfields Avenue, Wollongong, NSW 2522, Australia}
\author{P. Mikheenko}
\affiliation{Department of Physics, University of Oslo, P. O. Box 1048 Blindern, 0316 Oslo, Norway}

\date{\today}%

\begin{abstract}

\noindent Scaling behaviour of dynamically driven vortex avalanches in superconducting YBa$_{2}$Cu$_{3}$O$_{7-\delta}$ films deposited on tilted crystalline substrates has been observed using quantitative magneto-optical imaging. Two films with different tilt angles are characterized by the probability distributions of avalanche size in terms of the number of moving vortices. It is found in both samples that these distributions follow power-laws over up to three decades, and have exponents ranging between 1.0 and 1.4. The distributions also show clear finite-size scaling, when the system size is defined by the depth of the flux penetration front -- a signature of self-organized criticality. A scaling relation between the avalanche size exponent and the fractal dimension, previously derived theoretically from conservation of the number of magnetic vortices in the stationary state and shown in numerical simulations, is here shown to be satisfied also experimentally.

\end{abstract}

\maketitle

\section{Introduction}

\noindent Avalanche behaviour is found in a wide range of natural systems, and is commonly observed, e.g., as abrupt displacement in granular media, earthquakes, Barkhausen noise caused by sudden motion of magnetic domain walls in ferromagnetic materials, and abrupt displacement of vortices in superconductors, where their dissipative motion can trigger thermomagnetic avalanches \cite{Jensen, Pruessnerbook, Altshuler, Lightning}. Such avalanche activity is in general highly unwanted in practical applications of superconductors. However, its origin, which lies in the competition between intervortex interactions and quenched disorder, makes vortex matter an interesting system for studies of non-equilibrium dynamics.

Power laws in avalanche size, avalanche duration and temporal power spectra, together with finite-size scaling, are often interpreted as manifestations of self-organized criticality (SOC). This framework was originally developed from studies of a cellular automaton model of a sandpile \cite{BTW}. Although SOC seems to capture basic aspects of sandpile physics, both particle rolling and inertial effects tend to create system-spanning avalanches, which introduce cutoffs in the power-law probability distributions already after approximately one decade \cite{Jensen}. Such cutoffs are seen also in results obtained by numerical simulations \cite{Prado}. In contrast, studies of piles of rice grains have demonstrated power-law probability behaviour over nearly four decades \cite{Frette96, Aegerter2, Lorincz}. This increase was explained to be a result of rice grains having elongated shapes, and also being lighter, thus reducing both rolling and inertial effects.

In type-II superconductors an applied magnetic field is penetrating the material in the form of vortices. These flux lines have zero inertia and each carries one quantum of magnetic flux, $\Phi_0 = h/2e$. Here, $h$ is Planck's constant and $e$ the elementary charge. Moreover, also rolling effects are here nonexistent. Thus, the vortex matter represents a highly favourable system for studies of avalanche dynamics and SOC.

The vortices are affected by the Lorentz force from electrical currents flowing in the superconductor. However, this force can be counterbalanced by forces from microscopic material defects, which then serve as pinning sites for the vortex motion. Thus, in a slab geometry, vortices build up a metastable state \cite{Bean64} with a density profile quite analogous to the shape of a sandpile \cite{deGennes}. Just as a sandpile with constant slope is uniquely characterized by its angle of repose, the flux pile in a slab geometry has a constant flux density gradient, which by a Maxwell equation implies an electrical current flow of constant magnitude, the so-called critical current density, $j_c$. For thin films in perpendicular magnetic fields, the analogy is less obvious, \cite{Brandt93, Zeldov} since there the flux density gradient is not constant, although the critical state is still characterized by $j_c$.

Flux avalanches have been reported to occur in a wide variety of superconductors including Nb, Pb, Nb$_3$Sn, NbN, MgB$_2$, YNi$_2$B$_2$C, MoGe, MoSi and YBa$_{2}$Cu$_{3}$O$_{7-\delta}$ (YBCO) \cite{Duran95, menghini, Rudnev03, Rudnev05, Metallayer, MgbShantsev, Wimbush, Motta, Colauto, Leiderer, Aegerter, Qviller, Baziljevich}. The avalanches may be of thermomagnetic origin \cite{Rakhmanov}, or be caused by dynamically driven vortex rearrangements \cite{Altshuler, Qviller}. Both kinds of avalanches can occur in the same sample \cite{VlaskoVlasov, Behnia}. There is extensive experimental and numerical evidence that under some conditions, quantities measuring the size of dynamically driven vortex avalanches are distributed according to power-laws, and may also exibit finite-size scaling \cite{Altshuler, Aegerter, Qviller, Bassler, Welling}. For a recent review of SOC experiments in granular media and superconductors, see ref. \cite{Pruessner}.

In the present work, quantitative magneto-optical imaging (MOI) was used to measure the probability distributions of avalanche size in terms of number of vortices involved in two YBCO films deposited on tilted substrates with different tilt angles. It is shown that the distributions of avalanche size in terms of number of vortices follow power-laws. Moreover, for both samples, these distributions obey finite-size scaling. The avalanche size exponent and the fractal dimension are experimentally found to satisfy a scaling relation.

\section{Experimental}

When YBCO is deposited epitaxially on slightly tilted (vicinal) substrates, planar defects in the form of anti-phase boundaries are introduced with a period of 2-5 nm \cite{Jooss}. For tilt angles of $\theta \approx 10^{\circ}$, grain alignment and current-carrying abilities of such films are improved, thus making them interesting for  applications \cite{Jooss, Polyanskii1}. Such films have also potential use in Josephson junction circuits \cite{KimYoum, Mozhaev1, Mozhaev2}. YBCO films on tilted substrates have also other extended planar defects due to lattice mismatch. This results in anisotropic flux penetration and different critical current densities in the directions parallel, and perpendicular to these defects. The extended defects facilitate easy vortex motion \cite{Jooss, Polyanskii1, Djupmyr1}, and at low temperatures, the flux penetration becomes strongly quasi-one-dimensional by forming straight narrow channels.

For the present work, one film shaped as a strip of 0.9 mm width was deposited by laser ablation on a NdGaO$_3$ substrate with a tilt angle $\theta = 14\,^{\circ}$. Another film shaped as a square of side 4 mm was deposited by spin coating on a LaAlO$_3$ substrate with a tilt angle of $\theta = 20\,^{\circ}$. Both films are 200 nm thick and have critical temperatures, $T_c$'s, of 88 K and 90 K, respectively. For more sample preparation details, see \cite{Mozhaev2} and \cite{Qureishy}, respectively. The $\theta = 20\,^{\circ}$ sample was prepared like ``sample B" in ref. \cite{Qureishy}.

The samples described above have different $j_c$'s, and thus, slightly different experimental procedures were applied to investigate their flux dynamics. All the samples were initially zero-field-cooled (zfc) to $T = 4$ K.

The $14\,^{\circ}$-sample was then subjected to a field ramp to $B_a$ = 17.0 mT in equal 400 steps. The $20\,^{\circ}$-sample was subjected to a field ramp to $B_a = 8.5$ mT in 200 steps, thus, the same step $\Delta B_a = 42.5$ $\mu$T was used for both samples. At each new field, 5 magneto-optical images were recorded, after waiting 5 seconds for the vortex matter to relax. Subsequently, all groups of 5 images were averaged in order to reduce noise.

Before starting a new field ramp, $B_a$ was set to zero, and the temperature raised above $T_{c}$ to recreate a virgin state. In total, image series from 10 ramps were collected for the $14\,^{\circ}$-sample, and from 20 ramps for the $20\,^{\circ}$-sample.

MOI was performed using an in-plane magnetized ferrite-garnet film as Faraday rotating sensor \cite{Helseth, Goa}. This technique allows real-time visualization of flux distributions in superconductors, and allows high spatial ($\mu$m) and temporal (picosecond) resolution. Moreover, by MOI the entire flux pile can be observed, and thus, the statistics of internal avalanches and their morphology be characterized in great detail. The Faraday-rotating sensor film has a non-linear response to magnetic field and the lamp illumination in the setup is not fully uniform. Thus, to calibrate the image series, an additional field ramp from zero to 17.0 mT was performed at a temperature a few degrees above $T_{c}$. This allowed us to calibrate the magneto-optical response of the system, the measured raw light intensity $I(x,y)$ was fitted by a polynomial in $B_a(x,y)$ for every pixel of the CCD chip of the camera.

\section{Results}

\begin{figure}[t]
\includegraphics[width=20pc]{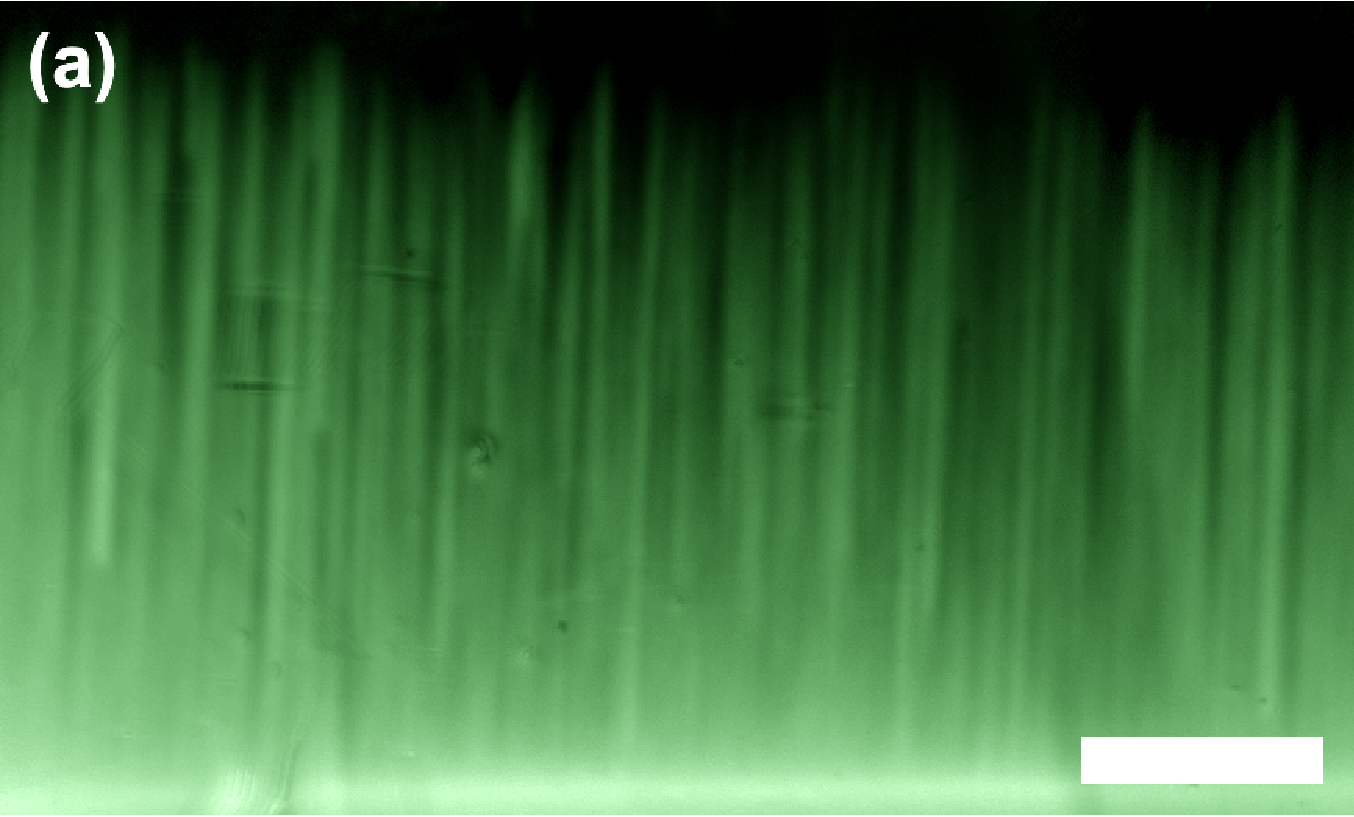}
\includegraphics[width=20pc]{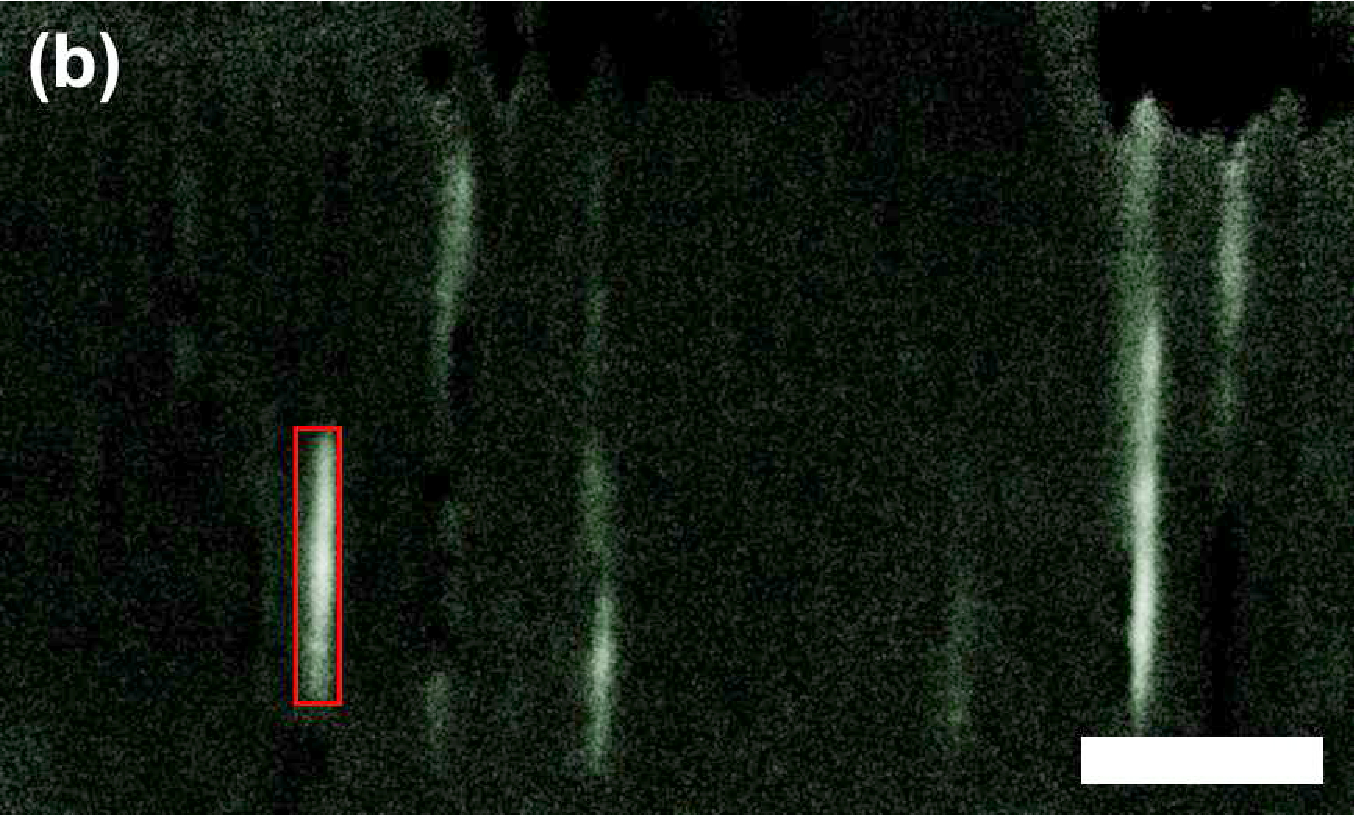}
\caption{\label{f1} (a) Magneto-optical image of flux penetration in the $14\,^{\circ}$-sample at $T = 4$ K and $B_a = 15.0$ mT. (b) Differential image at the same temperature and field as in (a) with $\Delta B_a = 42.5$ $\mu$T. An avalanche of length 170 $\mu$m and size 1280 $\Phi_0$ is marked with the red box. All scale bars are 150 $\mu$m long.}
\end{figure}

Presented in Fig. \ref{f1} (a) is a magneto-optical image of the flux density distribution near the long edge of the $14\,^{\circ}$-sample at $T = 4$ K. The edge itself is seen here as the bright horizontal line, since the external magnetic field piles up along the rim of the diamagnetic sample. One sees here a filamentary pattern of easy flux penetration. This pattern is due to reduced pinning along the extended defects in the superconducting sample caused by the substrate tilt.

At low temperatures, flux penetration in both samples is strongly intermittent in the form of quasi-1D avalanches along the ``channels". Seen in Fig. \ref{f1} (b) is a differential image obtained by subtracting the image in panel (a) from the next image in the series recorded during the field ramp. Thus, it shows the change in the flux penetration pattern after increasing the applied field by $\Delta B_a$. The differential image reveals that the flux penetration progresses in the form of quasi-one-dimensional avalanches along channels.

To analyze the avalanche activity, a computer program was used to extract quantitative parameters characterizing the statistical distribution of individual avalanche events. For that, the collected series of differential images were segmented into 10 intervals of $B_a$. This gives 40 differential images for the $14\,^{\circ}$-sample, and 20 differential images for the $20\,^{\circ}$-sample. In each interval, probability distributions of avalanche size in terms of magnetic flux were extracted. For both samples, a total of 400 differential images per series were available for statistical analysis. To ensure that a proper critical state was formed before the measurements, the first 120 images were discarded for the $14\,^{\circ}$-sample and the first 80 for the $20\,^{\circ}$-sample in every ramp.

In the differential images, a threshold value of image brightness was used to separate flux avalanches from background. Then, a median filter was applied to remove noise. The same threshold brightness and median filter were used for both samples, resulting in directly comparable probability distributions for the avalanche size in terms of amount of magnetic flux. The avalanche size is from now on defined as the number of vortices involved in a given avalanche event. This number was obtained by dividing the amount of flux in the avalanche by the flux quantum $\Phi_0 = 2.07 \cdot 10^{-15}$ Wb. The absolute frequencies of these parameters were binned in histograms, and each bin was subsequently divided by the total number of identified avalanches to yield probability distributions.

The probability distribution of avalanche size for the $14\,^{\circ}$-sample is shown in Fig. \ref{f2}, while the corresponding distribution for the $20\,^{\circ}$-sample is presented in Fig. \ref{f3}. The different colour-coded graphs correspond to different intervals of $B_a$, as indicated in the figures. Power-law behaviour in the avalanche size distribution is visible over about three decades for both samples. However, power-law behaviour is more clearly visible in Fig. \ref{f2} than in Fig. \ref{f3}. The smallest avalanches included in these data consist of about 10 vortices, whereas the largest are rearrangements of 10000 vortices or more.

\section{Data analysis}

\begin{figure}[t!]
\includegraphics[width=19pc]{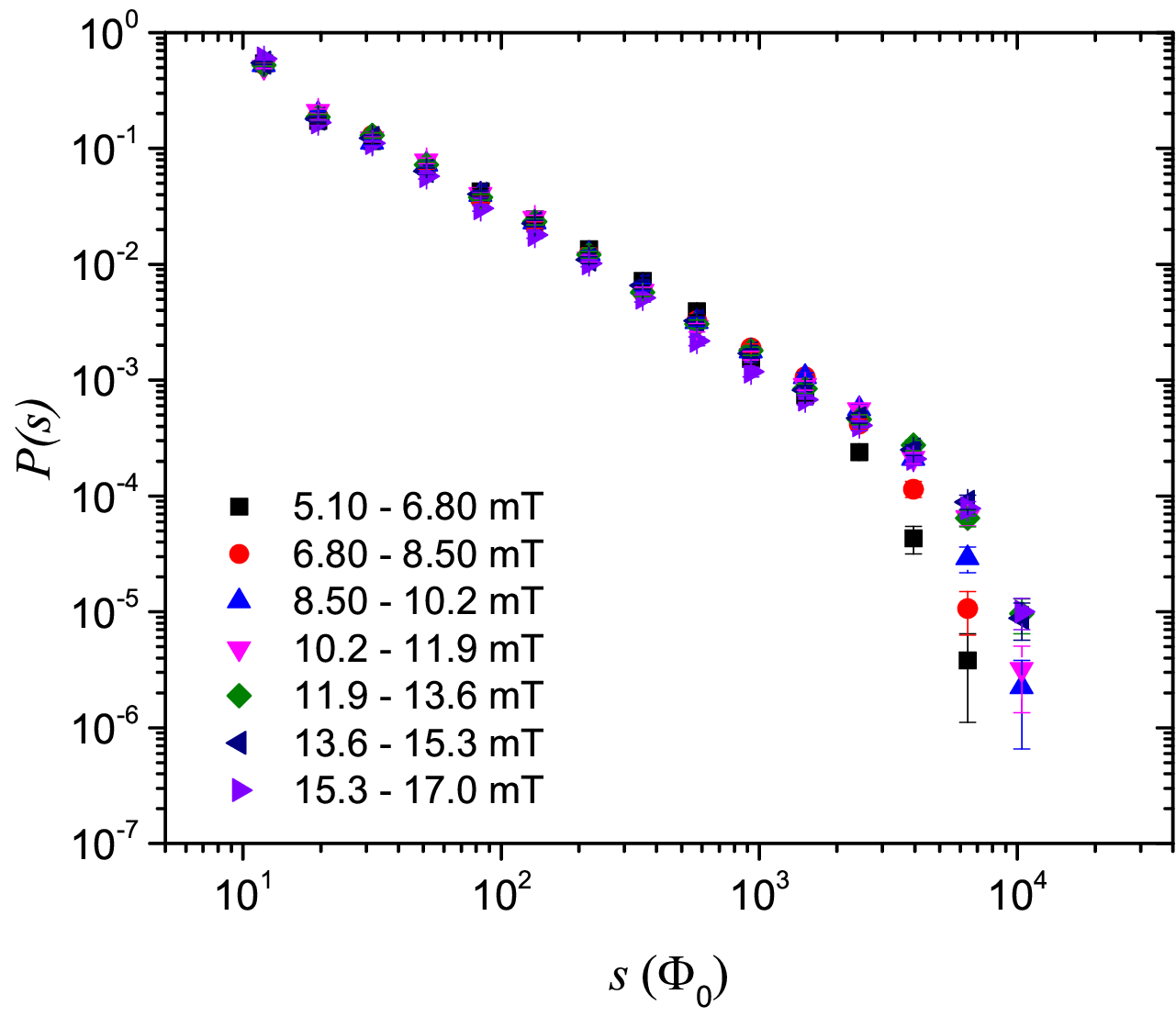}
\caption{\label{f2} Probability distributions of avalanche size $s$ in the $14\,^{\circ}$-sample.}
\end{figure}

\begin{figure}[t!]
\includegraphics[width=19pc]{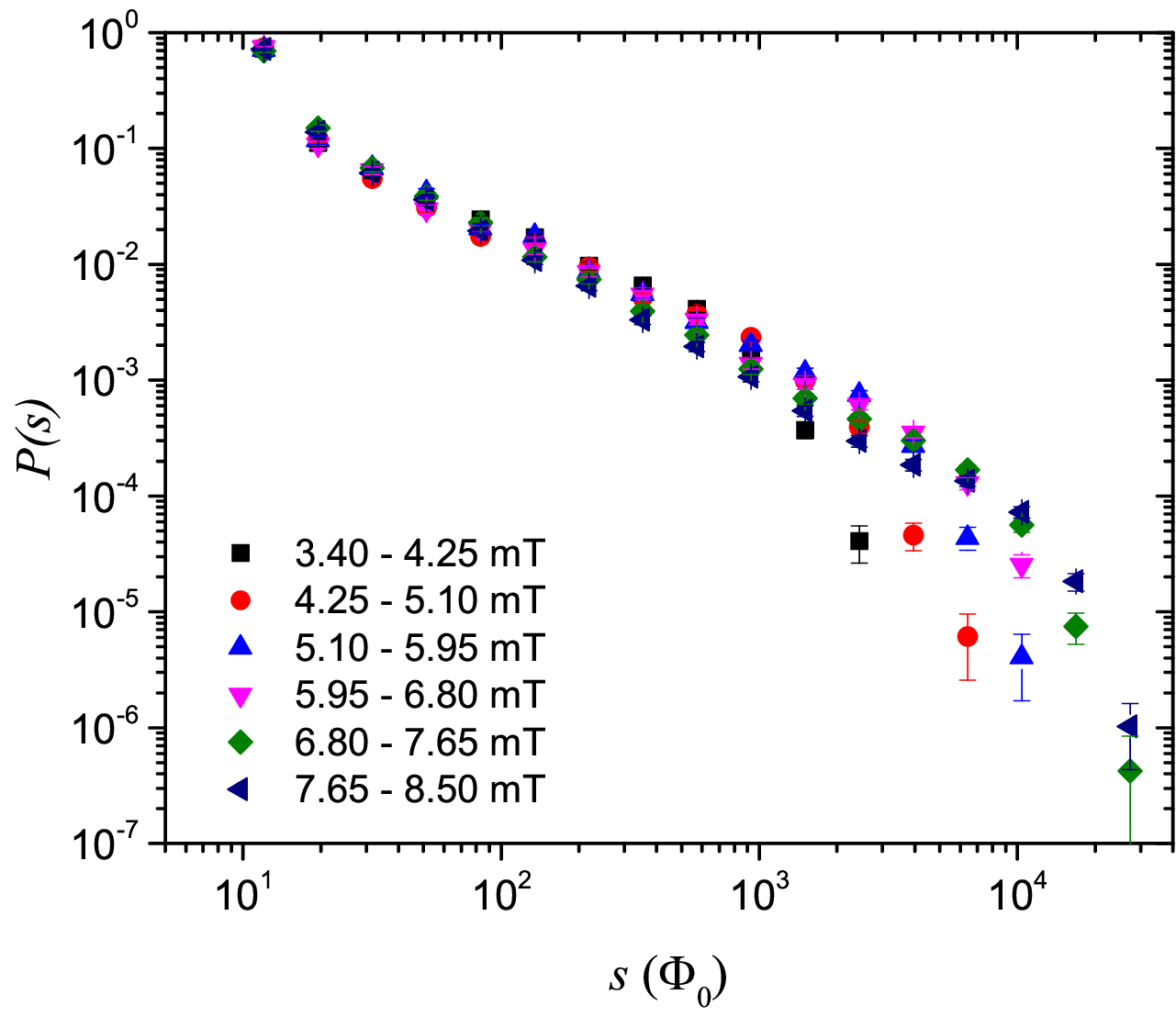}
\caption{\label{f3} Probability distributions of avalanche size $s$ in the $20\,^{\circ}$-sample.}
\end{figure}

The distributions shown in Figs. \ref{f2} and \ref{f3} look like approximate power-laws with cutoffs in the system size $L$. Here, $L$ is the penetration depth of the flux front at the applied field halfway between the start and end values of the intervals of $B_a$. The penetration depth $L$ is manually measured from the recorded images.

We will now investigate in more detail whether functions on the form of power-laws with cutoffs $\mathcal{P}(s, L)$ can describe the observed behaviour of the distributions of avalanche size $s$. Consider therefore functions of the form

\begin{equation}
 \mathcal{P} (s, L) \propto s^{-\tau} \,  f \!
\left(\frac{s}{L^D}\right).
\label{eq1}
\end{equation}

Here, $\tau$ is the avalanche exponent for the size distribution. The scaling function $f$ is constant up to a cutoff scale and thereafter falls off as a function of $s/L^D$. In order to check whether the distributions obey finite-size scaling, we start by rearranging Eq. \ref{eq1}, so that $f$ is isolated on the right-hand side. E.g., we plot $s^{\tau} \mathcal{P}(s, L)$ against $s/L^D$ to investigate the functional form of $f$. If the distributions now collapse onto one curve when this is done for all the intervals of $B_a$, finite-size scaling is present. Moreover, the better the curve collapse, the more exact is the scaling \cite{Aegerter}.

Furthermore, the avalanche exponent and fractal dimension are determined by adjusting them until optimal curve collapse is obtained. A plateau will appear in the collapsed curves when the avalanche exponent is optimally chosen. The plateau corresponds to the power-law region between the lower and upper cutoffs in the distributions. The fractal dimension, on the other hand, is obtained by adjustment until the upper cutoffs in the distributions are best aligned. This was done for both probability distributions measured in this work, and the results are summarized in Table \ref{t1}.

\begin{table}[h]
\centering
\caption{Avalanche exponents $\tau$ and fractal dimensions $D$.}
\begin{tabular}{|l|l|l|l|l|} \hline
                    Sample & Avalanche exponent  & Fractal dimension    \\ \hline
           $14\,^{\circ}$  & $1.30$ $(\pm0.03)$  & $1.43$ $(\pm0.08)$   \\ \hline
           $20\,^{\circ}$  & $1.06$ $(\pm0.03)$  & $1.40$ $(\pm0.10)$   \\ \hline
\end{tabular}
\label{t1}
\end{table}

\begin{figure}[t!]
\includegraphics[width=19.5pc]{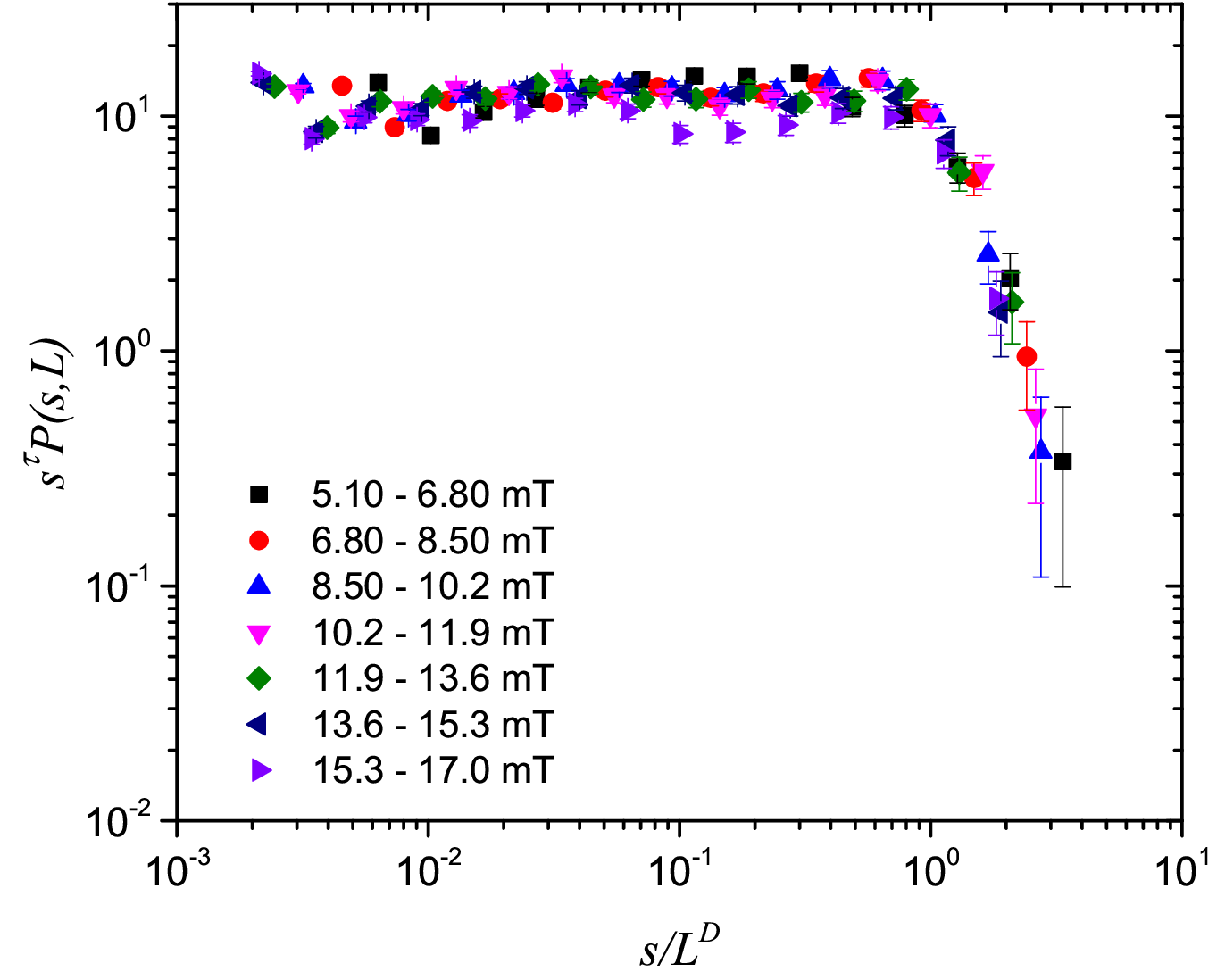}
\caption{\label{f4} Finite-size scaling of avalanche size probability distributions $\mathcal{P} (s, L)$ in the $14\,^{\circ}$-sample.}
\end{figure}

\begin{figure}[t!]
\includegraphics[width=19.5pc]{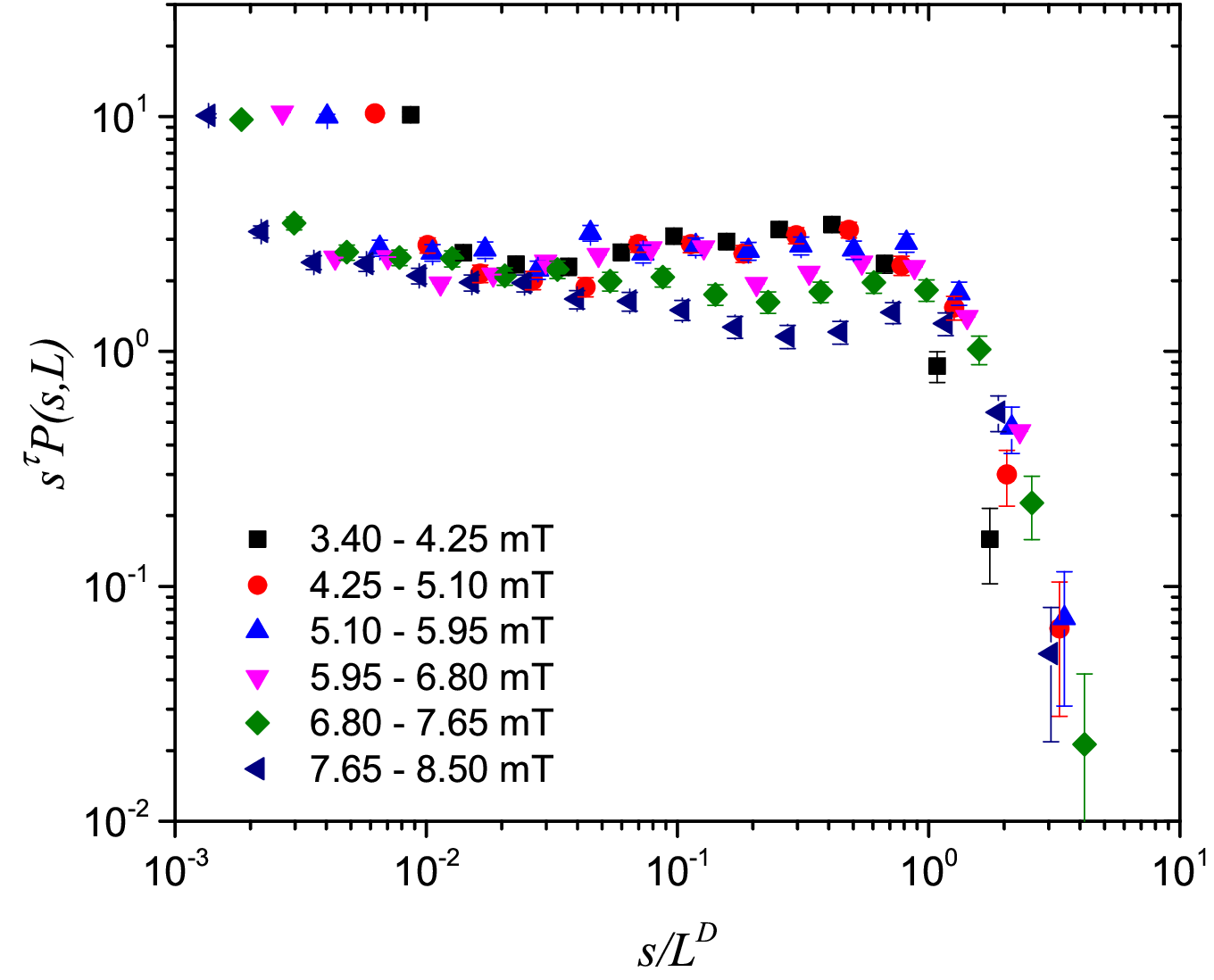}
\caption{\label{f5} Finite-size scaling of avalanche size probability distributions $\mathcal{P} (s, L)$ in the $20\,^{\circ}$-sample.}
\end{figure}

Shown in Figs. \ref{f4} - \ref{f5} are curve collapses of the distributions in Figs. \ref{f2} - \ref{f3}, respectively. Both curve collapses show expected deviations from power-law behaviour and scaling near the lower cutoff. It is clear from Figs. \ref{f4} - \ref{f5} that finite-size scaling is more pronounced in the data from the $14\,^{\circ}$-sample than for the $20\,^{\circ}$-sample.

We now proceed to the central result of the present work, namely, the experimental verification of the scaling relation Eq. $\ref{eq3}$

\begin{equation}
 D(2-\tau) = 1
 \label{eq3}
\end{equation}

This equation was derived from the conservation of the number of magnetic vortices in the stationary state in \cite{Bassler}, and also shown in numerical simulations in the same work. It has later also been demonstrated numerically in ref. \cite{Cruz}. A statement on its validity in the experimental context of the present work is given in \cite{Pruessnernote}.

Consider now the correspondence between the experimentally obtained values in Table \ref{t1} and the scaling relation, Eq. $\ref{eq3}$. For the $14\,^{\circ}$-sample, good curve collapses are obtained with $D$ in the range $1.35$ to $1.51$ and with $\tau$ in the range $1.27$ to $1.33$. Inserting these values of $D$ and $\tau$ in Eq. $\ref{eq3}$
shows that Eq. $\ref{eq3}$ is experimentally satisfied within 10 \%.

The analysis was repeated for the $20\,^{\circ}$-sample, where the result is $D(2-\tau)$ in the range 1.18 to 1.46. The reduced agreement here with Eq. $\ref{eq3}$ is most likely because this avalanche size distribution shows deviations from a pure power-law, which is a prerequisite for obtaining Eq.~$\ref{eq3}$.

\section{Discussion}

From the above analysis, it is clear that in both samples, the probability distributions of avalanche size demonstrate power-law behaviour and finite-size scaling consistent with SOC. The distributions show these behaviours over two to nearly three decades. However, the power-laws and scaling is more evident in the $14\,^{\circ}$-sample than in the $20\,^{\circ}$-sample. Molecular dynamics simulations have shown a similar breakdown to ``dirty power-laws" when length scales, in addition to the system size, are introduced by low density of pinning sites and the resulting formation of ``vortex rivers" \cite{Olson}. However, determination of the precise origin of the deviations is beyond the scope of this work.

Similar results obtaind by MOI are presented in the work \cite{Aegerter}, reporting $\tau = 1.29\pm0.02$ and $D = 1.89\pm0.03$, and \cite{Welling}, where it was found that $\tau = 1.07\pm0.02$ and $D = 2.25\pm0.05$. The obtained avalanche size exponents are in the same range as those found in the present work, but the obtained fractal dimensions differ considerably. The experiment of Field et al. \cite{Field}, utilizing a hollow-cylinder geometry, found avalanche size exponents in the range $\tau = 1.4-2.2$. However, the exponents found in such off-edge avalanches are not necessarily directly comparable with internal avalanches \cite{Pruessner}, which are the type considered in the present work. Exponents differing considerably from those observed in the present work were found in \cite{Behnia} with $\tau = 2.05$ and \cite{Altshuler3} with $\tau = 3.0$. However, those experiments had single-vortex resolution and probed a different range of the probability distribution functions than that covered by MOI experiments.

Molecular dynamics simulations \cite{Olson} gave an exponent in the range of $\tau = 0.9-1.4$ for the number of displaced vortices (avalanche size) in systems with high pinning density. The exponent was found to increase with higher pinning strength. The avalanche size exponents found in the present work are also not very different from those obtained from simulations of the Bassler-Paczuski model, a cellular automaton for 2D flux penetration \cite{Bassler}. The original work by Bassler and Paczuski gave $\tau = 1.63\pm0.02$ and $D=2.75\pm0.1$. A newer work by Cruz et. al. \cite{Cruz} on a variation of this model with strong, periodic and densely spaced pinning sites resulted in $\tau = 1.45\pm0.02$ and $D=2.2\pm0.1$. The Bassler-Paczuski model shows robust SOC behaviour over four decades for a variety of different parameters in the pinning landscape \cite{Bassler, Cruz}. However, both the molecular dynamics simulations and the cellular automata models assume short-ranged vortex-vortex interactions for the bulk case, whereas in thin films the vortex-vortex interactions fall off as $1/r$, and probably lead to results in a different universality class. Note that the exponent and fractal dimension from the work of \cite{Bassler} is in excellent agreement with Eq. \ref{eq3}. Moreover, the agreement is also quite good when the exponent and fractal dimension from the work of Cruz et. al. \cite{Cruz} is used.

\section{Conclusions}

In summary, the probability distributions of flux avalanches were measured by quantitative MOI for two YBCO samples deposited on substrates cut with tilt angles of $14\,^{\circ}$ and $20\,^{\circ}$. Probability distributions of avalanche size in terms of numbers of vortices were extracted from the data. These distributions follow approximate power-laws over up to three decades, and demonstrate finite-size scaling. Avalanche exponents and fractal dimensions were obtained by careful inspection of the finite-size scaling curves. The obtained exponents are between 1.0 and 1.4, and the avalanche size exponents determined in this work are similar to those that has been found in other MOI experiments on superconductors, molecular dynamics simulations and cellular automata models. The scaling relation $D(2-\tau) = 1$ between the avalanche size exponent and the fractal dimension, previously derived theoretically from conservation of the number of magnetic vortices in the stationary state and shown to be satisfied in numerical simulations, was also experimentally proven with an accuracy of 10\% for the $14\,^{\circ}$ sample.

\section{Acknowledgments}

The Authors want to thank Julia E. Mozhaeva for her kind assistance during sample preparation. This work was financially supported by the National Natural Science Foundation of China (51571002) and Beijing Natural Science Foundation (2172008). The work performed at IPT RAS was supported by the State Program of the Russian Ministry of Science and Education. This is a pre-print of an article published in Scientific Reports. The final authenticated version is available online at: https://doi.org/10.1038/s41598-020-62601-y

\end{document}